\documentclass[11pt]{article}

\usepackage[english]{babel}
\usepackage{fullpage,amssymb,amsmath,amsthm,graphicx}
\usepackage{epsf}
\usepackage{epsfig}
\usepackage{setspace}
\usepackage{multirow}
\usepackage{subfigure}
\usepackage{comment}

\usepackage{graphics}
\usepackage{graphicx}
\usepackage{geometry}
\usepackage{array}
\usepackage{float}
\usepackage{lscape}
\usepackage{color}
\usepackage{calligra}
\usepackage{dsfont}
\usepackage{tabularx}
\usepackage{dcolumn}
\usepackage[bottom]{footmisc}
\usepackage{enumerate}
\usepackage{url}

\newtheorem{proposition}{Proposition}
\newtheorem{corollary}{Corollary}

\theoremstyle{remark}

\begin{document}

\title{On-Chain Auctions with Deposits}

\author{Jan Christoph Schlegel \thanks{\textit{Corresponding author}: Department of Economics, City, University of London, Northampton Square, London EC1V 0HB, United Kingdom, e-mail:
\texttt{jansc@alumni.ethz.ch}} \and Akaki Mamageishvili
\thanks{ETH Zurich, Department of Management, Technology and Economics, Zürichbergstrasse 18, 8092 Zürich, Switzerland}}

\maketitle
\begin{abstract}
  Sealed-bid auctions with deposits are frequently used in blockchain environments. An auction takes place on-chain: bidders deposit an amount that fully covers their bid (but possibly exceeds it) in a smart contract. The deposit is used as insurance against bidders not honoring their bid if they win. The deposit, but not the bid, is publicly observed during the bidding phase of the auction.
  
  The visibility of deposits can fundamentally change the strategic structure of the auction if bidding happens sequentially:  Bidding is costly since deposits are costly to make. Thus, deposits can be used as a costly signal for a high valuation. This is the source of multiple inefficiencies: to engage in costly signalling, a bidder who bids first and has a high valuation will generally over-deposit in equilibrium, that is, deposit more than he will bid. If high valuations are likely there can, moreover, be entry deterrence through high deposits: a bidder who bids first can deter subsequent bidders from entering the auction. Partial pooling can happen in equilibrium, where bidders of different valuations deposit the same amount. The auction fails to allocate the item to the bidder with the highest valuation.
\end{abstract}

\section{Introduction}
On-chain implementations of auctions often take the following form: bidders deposit an amount that fully covers their bid in a smart contract. The deposit is used as insurance against bidders not honoring their bid if they win. The deposit, but not the bid, is publicly observed during the bidding phase of the auction. In the case of a second-price auction,~\cite{VCG}, after bidding is closed the winner is determined by evaluating the non-public bids. The highest bid wins and the winner pays the second highest bid. Leftover deposits are repaid. 

\begin{figure}[t]
 \includegraphics[width=0.75\textwidth]{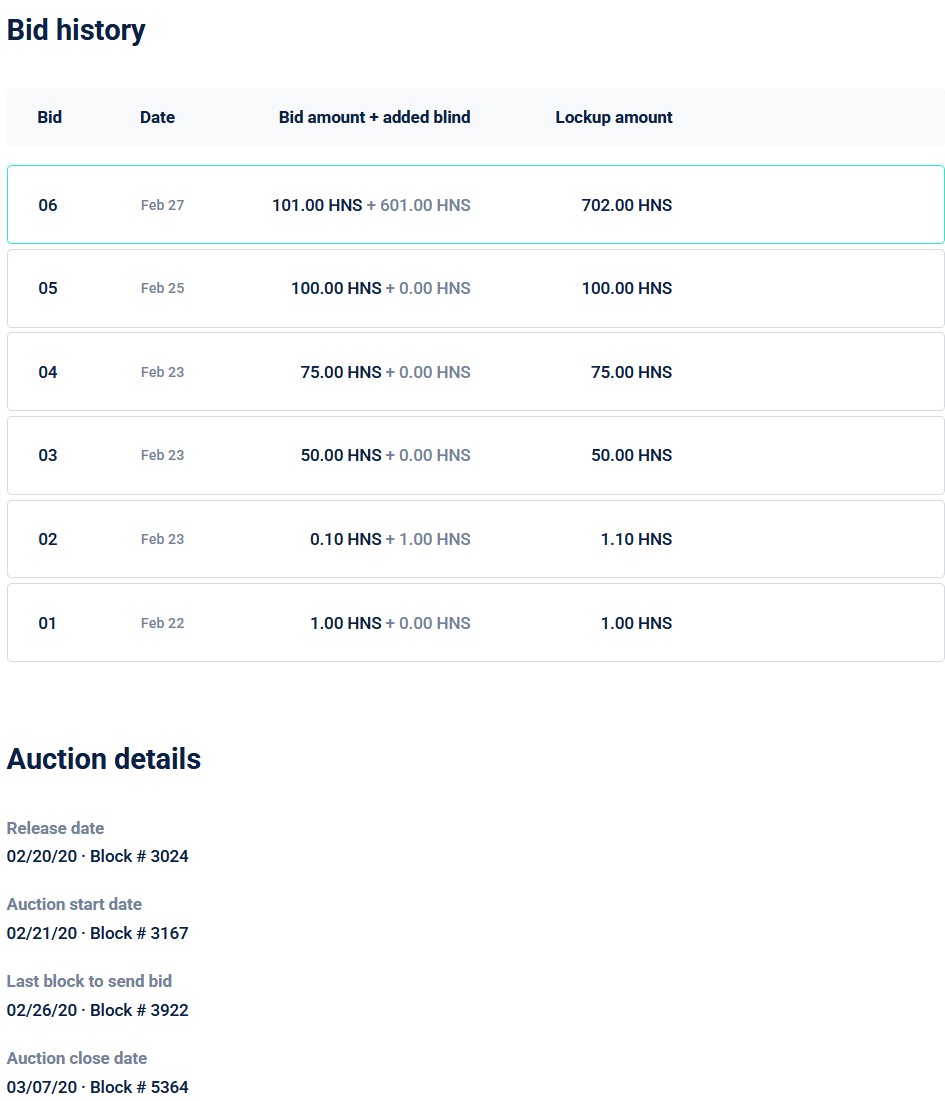}
 \caption{Bid history and auction details for the top level domain ".scholar" on namebase.io.}
 \label{scholar}
 \end{figure}

Often it is claimed that this auction format is equivalent to a standard second-price auction without depositing. While this  is indeed a good approximation if the value of the auctioned item is relatively small for all bidders, the auction has fundamentally different properties if the value of the item is substantial for many bidders. In particular, we claim that this auction format can be vulnerable to entry-deterrence strategies. To illustrate this point consider the instance of the top-level domain (TLD) auction on "namebase.io" for the TLD ".scholar" as reproduced in~Figure~\ref{scholar}: namebase.io is an example of a marketplace that uses the auction format that we analyze in this paper. Bidders can bid on a TLD such as ".scholar" in the example in Figure~\ref{scholar}. Bidders submit a bid and a blind to the auction. The sum of the two numbers (the "lockup amount"), but not the bid itself, is visible to other bidders during the bidding phase of the auction. At the end of the bidding phase, bids are evaluated according to a second price auction and left-over lockup amounts are repaid.  As can be observed in the bid history reproduced in Figure~\ref{scholar}, the sixth bid in the auction involves a substantial lockup amount. Subsequently, no further bid is made. Note that the sixth bid involves substantial over-depositing, i.e.~the blind part is much larger than the actual bid. The bidder seems to have successfully engaged in entry deterrence: the large lockup amount likely discouraged further potential bidders to enter the auction and place a bid. In this paper, we want to give a theoretical explanation of this kind of bidding behavior.

The logic of our analysis is as follows: Bidding is costly since depositing is costly. If the deposit has to be staked for a substantial amount of time\footnote{For namebase's auction deposits are kept for 10 days after the end of the bidding phase.} 
and can not be used or sold meanwhile, the bidder is, for example, exposed to the volatility of the underlying token price. Other costs of depositing can come from transaction costs when buying the native token that is used to make the deposit, from borrowing costs or from transactions cost for processing the bid on-chain. Even if this cost is relatively small, it can have substantial strategic implications if valuations for an item are expected to be high\footnote{Bids can be substantial in these kinds of auctions. 
For the domain auctions organized by namebase.io, domains have been sold for close to 100000 $\$$: \url{https://newssoso.com/2021/02/03/this-nft-web-domain-just-sold-for-record-breaking-84000/}. Such auctions can also suffer from speculation, since they allow instant resale,~\cite{resale}.}. A bidder can deter entry into the auction by depositing a large amount. This is a costly signal to other bidders that the bidder likely has a high valuation. Another bidder wondering whether to enter the auction infers that this makes it unlikely that he wins the auction, and that he is likely to pay a high price if he wins the auction. Since bidding is costly and the cost of bidding increases in the size of the deposit, he possibly refrains from bidding at all. Since bids are not public, this could be even in the case when he would have had a higher valuation than the first bidder, but the first bidder successfully deters him through over-depositing. Efficiency is lost: with positive probability the item does not go to the bidder with the highest willingness to pay. Moreover, revenue is severely decreased since bidders are deterred from entering the auction. Finally, wasteful over-depositing will take place, bidder $1$ will deposit a large amount exceeding his true valuation amount to send a costly signal of his high valuation.

Interestingly the strategy outlined here goes in the opposite direction of other strategies usually observed in online auctions: bidders can have an advantage if they move early because they can scare away other bidders by over-depositing, whereas in other environments bidders want to get late into the auction in order to ''snipe"~\cite{roth2002}.

We flesh out this logic in a stylized model.  
We study the case of sequential bidding with two bidders. If bidder 1 bids before bidder 2, bidder 1's deposit is visible to bidder 2. This drastically changes the strategic interaction and the equilibrium bidding behavior.
Generically, the first bidder will over-deposit if he has a high valuation. This over-depositing is wasteful since it is costly to the bidder and only satisfies the purpose of signalling a high valuation. In addition to the efficiency loss due to over-depositing, there can be additional allocative inefficiencies due to entry deterrence which can come in two forms:
\begin{enumerate}
    \item Unless there is a very high likelihood of low valuations for the second bidder, the first bidder will refrain from entering the auction if he has a low valuation since he wants to save on the depositing cost when facing a low likelihood of winning the auction. 
    \item More severely, if high valuations are likely, equilibria exists in which high types of the first bidder  are pooled. In the most extreme case, there can exist equilibria with just two deposit levels. The first bidder will choose either not to bid at all or to over-deposit. If the second bidder observes a large deposit he can be deterred from entering the auction since he assigns a substantial likelihood to the first bidder having a high valuation. Thus, he will not bid at all unless his valuation is very high. The equilibrium can be very wasteful. The first bidder can win the auction even though he could have a substantially lower valuation for the item than the second bidder. 
\end{enumerate}





As a benchmark, in the appendix, we consider the case of simultaneous depositing or equivalently the case where privacy-preserving depositing is available so that no information is revealed to other bidders. In this case, bidding in equilibrium is approximately truthful. Bids are only shaded by a small amount to incorporate depositing costs. Social welfare is approximately maximised since the item is allocated to the bidder with the highest valuation, and the only loss in welfare is due to depositing cost.

\subsection{Empirical Evidence}
Among entries in namebase.io's domain database we have calculated several summary statistics for the subset of  around $5.9$ million domains which were resold. Here we take resale as in indicator of relevance. 
First, we note that the number of bidders for each domain is generically low. Figure~\ref{fig:my_label} shows the skewness of the distribution. While the highest number of bidders for one domain is $28$, domains with at most $2$ bidders comprise more than $75\%$ and domains with at most $4$ bidders comprise more than $96\%$ of all domains. These observations are in line with our assumption on having only $2$ bidders. Over-depositing happens in $58\%$ of all bids. In $61\%$ of the sold domains, the winner bidder overdeposits. In $55\%$ of the cases, the winner bidder over-deposits and is also the last bidder. Last two observations can be interpreted as an evidence that overdepositing serves the purpose of deterring other bidders to enter bidding process. In Figure~\ref{overdepositing}, we plot the ratio between over-deposited quantity over bids, conditional that the bid size is at least $10$. That is, we condition on cases where the bids are significantly high. The histogram suggests that depositing incurs at least some opportunity cost, as high ratios diminish. On the other hand, it is still present as a mechanism of signalling.

\subsection{Literature}
 
Exploiting auctions through "front running" in the blockchain environment is studied in~\cite{flash_loan} and~\cite{hafner2021}. While the entry-deterrence strategies analyzed in our paper can be interpreted as a form of front-running, they are qualitatively very different. Whereas conventional front-running is used to exploit informational advantages, in our setting bidders want to go early to signal information to others. Sequential bidding with constant cost of entry and deterrence is a topic of~\cite{seq_costly_entry}. With constant cost of entry instead of a constant marginal cost of entry, bidders can only signal through their entry decision but not through deposits. We use standard terminology from signalling games~\cite{signalling} and related solution concepts~\cite{fudenberg}.  Costly signalling in auctions, although in a different environment, has been considered before, e.g.~in~\cite{horner2007}. Deposit costs in on-chain exchange mechanisms are the topic of~\cite{mamageishvili2020}. Our paper is remotely related to the literature on all-pay auctions,~\cite{all_pay_auctions}. In all pay auctions every bidder pays their bid completely, while in our model only some percentage of the bid is paid, because of the opportunity costs of depositing. Sunk costs or, equivalently, money burning can sometimes be useful in optimal mechanism design,~\cite{money_burning}, in particular, increasing expected surplus.

\section{Model and Results}\label{sec:sequential}
We consider two bidders with independently and identically distributed valuations $v_1,v_2\in (0,1)$ distributed according to a (continuously differentiable) distribution $F$ with density $f$.

Bidders submit a bid $b(v_i)$ and a deposit $d(v_i)$ such that $0\leq b(v_i)\leq d(v_i)$. Note that both are functions of the real valuation $v_i$. The constant marginal cost of depositing is $c>0$. The winner of the item is determined through a second price auction. That is, if $i$ makes the winning bid, equivalent to $b_i>b_j$, then his profit is
$$v_i-b(v_j)-cd(v_i),$$
whereas $j$ loses $$-cd(v_j).$$
Since we have a second price auction, it will generally be the case that in equilibrium a bidder either bids his valuation or his full deposit,$$b(v_i)=\min\{v_i,d(v_i)\}.$$

 In the following we consider sequential bidding. As a benchmark for comparison we consider the case of simultaneous bidding in Appendix~\ref{Simultanous}.
%
We assume that bidder $1$ moves first and bidder $2$ moves second so that bidder $1$'s deposit (but not his bid) is known to bidder $2$ before he chooses his bid.
We use the tie-breaking rule that the second bidder wins in case of equal bids.
As observed before, bidder $1$ bids either his deposit or his valuation so that the strategy of bidder $1$ is characterized by a deposit function $d_1:[0,1]\rightarrow [0,\bar{d}]$ where $\bar{d}\geq 1$ is the maximal deposit a bidder can make.\footnote{Typically, we think of $\bar{d}$ as a large quantity. It could for example be the number of tokens in the protocol in total.} It is always optimal for Bidder $1$ to bid $\min\{v_1,d_1(v_1)\}$.

For bidder $2$, it is no loss of generality to assume that he does not over-deposit. Otherwise, if $b_2<d_2$, then the bidder could deposit $b_2$ instead without changing his bid and become strictly better off by saving on depositing cost. So bidder $2$'s strategy is characterized by his deposit function  $d_2:[0,1]\times[0,\bar{d}]\rightarrow[0,1]$  where $d_2(v_2,d_1)$ is his choice of deposit under valuation $v_1$ if he observes bidder $1$ depositing $d_1$. It is always optimal for Bidder~2 to bid $b_2=d_2(v_2,d_1)$.

Sequential bidding turns our model into a signalling game. Given that bidders' strategies are fully determined by their depositing functions, we understand a {\it Perfect Bayesian Equilibrium} to be a triple $(d_1,d_2,\mu_2)$ of deposit functions for the two bidders and beliefs $\mu_2:[0,1]\times[0,\bar{d}]\rightarrow[0,1]$ for bidder $2$\footnote{We do not need to specify bidder~1'$s$ beliefs about bidder~$2$'s valuation as they will coincide with the prior $F$.} where $\mu_2(\cdot,d_1)$ is the cdf for $v_1$ conditional on observing a deposit $d_1$ 
such that 
\begin{itemize}
\item[1.] Each player's strategy specifies optimal actions, given his beliefs and the strategies of the other player, i.e.~ $d_1(v_1)$ is a solution to 
\begin{align*}
\max_{d_1} Pr_{v_2\sim F}[b_1>b_2](v_1-E_{v_2\sim F}[b_2|b_1>b_2])-cd_1,\\
\text{ where }b_1=\min\{v_1,d_1\},\quad b_2=\min\{v_2,d_2(v_2,d_1)\},
\end{align*}
and $d_2(v_2,d_1)$ is a solution to
\begin{align*}
\max_{d_2} Pr_{v_1\sim \mu_2(d_1)}[b_2\geq b_1](v_2-E_{v_1\sim \mu_2(d_1)}[b_1|b_2\geq b_1])-cd_2,\\
\text{ where }b_1=\min\{v_1,d_1\},\quad b_2=\min\{v_2,d_2\}.
\end{align*}

\item[2.] Beliefs are updated according to Bayes' rule whenever applicable, for $0\leq d_1\leq \bar{d}$ such that there is a type $v_1$ with $d_1(v_1)=d_1$ we have $$\mu_2(v,d_1)=F(v|d_1(v_1)=d_1).$$
\end{itemize}
 All subsequently constructed Perfect Bayesian equilibria will also satisfy standard refinement properties for signalling games like the Cho-Kreps intuitive criterion~\cite{cho1987}.
Following standard terminology of the signalling literature, we call an equilibrium {\it separating}, if different types send different messages. In our setting, the type of bidder $1$ is the valuation $v_1$, while a message is the deposit $d_1(v_1)$.
Therefore, in a separating equilibrium, the second bidder can exactly determine the valuation of the first bidder from the deposit.
We call an equilibrium {\it separating conditional on entry} if conditional on the first bidder depositing a positive amount, the second bidder can exactly determine the valuation of the first bidder from the deposit. Similarly, we call an equilibrium, {\it essentially separating (conditional on entry)} if it is separating (conditional on entry) for all types below $\frac{1}{1+c}$. We call an equilibrium {\it pooling equilibrium} with $n$ deposit levels if bidder~1 uses the same deposit under different valuations and $n$ different deposit levels are observed in equilibrium.

\subsection{Separating equilibria}
Suppose there is a separating equilibrium, $d_1(v_1)\neq d_1(v_1')$ for $v_1\neq v_1'.$ In that case, bidder $2$ knows bidder $1$'s valuation when bidding. The equilibrium depositing function will be increasing, $v_1>v_1'\Rightarrow d_1(v_1)>d_1(v_1').$ Let ${d}:=d_1(1).$ Let $u(d_1)$ be the inverse of the depositing function. For $0\leq d_1\leq d$ bidder $2$ will bid:

\begin{equation}
d_2(v_2,d_1)=\begin{cases}\min\{d_1,u(d_1)\},\text{ for }v_2\geq(1+ c)\min\{d_1,v(d_1)\}\\0,\text{ otherwise.}
\end{cases}
\end{equation}


Given the bidding function for bidder $2$, the expected utility of bidder $1$ when depositing $d_1$ is 

\begin{equation}\label{expected_utility}
    v_1F((1+c)\min\{d_1,u(d_1)\})-cd_1.
\end{equation}

In general, the depositing function $d_1$ as a function of valuation should satisfy the first order condition:
\begin{align}\label{FOC_gen_under}
(1+c)v_1f((1+c)d_1)=c, \quad \text{ for }d_1(v_1)\leq v_1,\\
(1+c)v_1f((1+c)v_1)=cd_1'(v_1), \quad \text{ for }d_1(v_1)> v_1.\label{FOC_gen_over}
\end{align}

The latter condition is obtained by differentiating the expected utility of the first bidder,~\eqref{expected_utility}, with respect to $u$, and evaluating it at $v_1$.\footnote{The first order condition in case of over-depositing resembles the condition that characterizes equilibrium bidding strategies in an all-pay auction with the difference that in our case the depositing function is scaled by the factor $c>0$. The same kind of condition would characterize bidding strategies in an auction format where all bidders pay a fraction $c$ of the sum of bids.} Subsequently, we use these first order condition to derive separating equilibria (resp.~equilibria that are separating conditional on entry) for different distributions.

\subsubsection*{Separation conditional on entry}
While fully seperating equilibria only can exist if the distributions is concave for low types (see below), we can obtain equilibria with separation conditional on entry for any unimodal distribution:
In the following, we call a distribution on $[0,1]$ unimodal, if there is a $0\leq m\leq 1$ such that $F$ is convex on $[0,m]$ and concave on $[m,1]$. Since we allow for the possibility of weak convexity and weak concavity in the definition (the uniform distribution is an example of a unimodal distribution according to this definition), $m$ does not need to be unique, and we use the notational convention that $m_F$ is the maximal such $m$ for distribution $F$ so that the distribution is strictly concave on $[m,1]$ (e.g. for the uniform distribution $m=1$).
For the "convex part" of the distribution, we obtain no-entry for low types and (potentially) over-depositing for higher types. For the "concave part" we can have over- as well as under-depositing.
\begin{proposition}\label{separationconditionalonentry}
For a unimodal distribution with $E[v]\geq c$, there exists a maximally efficient equilibrium which is essentially separating conditional on entry. Bidder $1$ under-deposits for low types and over-deposits for high types.
\end{proposition}

\subsubsection*{Separation for concave distributions}
The existence of a separating equilibrium where bidder $1$ always enters the auction requires that the density function $f$ is strictly decreasing in the beginning, i.e.~the cumulative distribution function is concave for low types and hence low valuations are more likely. This is the case, in particular if the entire distribution is concave in which case we obtain:






\begin{corollary}
\label{partially_separated}
For a strictly concave distribution, there exists an essentially separating equilibrium for which we have under-depositing for low types
\end{corollary}
\subsubsection*{Over-depositing for convex distributions}
Conversely to the previous result, if the density function is strictly increasing in the beginning, bidder $1$ will not enter the auction for low types. More generally, if the entire distribution is convex, conditional on entry we observe over-depositing:
\begin{corollary}
\label{uniform_separation}
For a convex distribution with $E[v]\geq c$, in equilibrium bidder $1$ does not enter the auction for low types. There exists an equilibrium which is essentially separating conditional on entry. In each such equilibrium, conditional on entry, bidder $1$ over-deposits.
\end{corollary}

\subsection{Pooling equilibrium with two deposit levels}
If high valuations are more likely than low valuations we can construct pooling equilibria. We consider the case of only two deposit levels in equilibrium,  bidder $1$ either deposits $0$ or the maximal valuation $1$, and show that such an equilibrium can be constructed for the quadratic distribution.

To construct the equilibrium, consider a marginal type of bidder $1$, $0<u<1$, such that he deposits $0$ if his valuation is below $u$ and $1$ if his valuation is above $u$.
In that case, if bidder $2$ observes a deposit of $0$, he will bid $0$ himself and win the object for free. If bidder $2$ observes a deposit of $1$, he knows that the valuation of bidder $1$ is between $u$ and $1$. We also need to specify what bidder $2$ believes off-equilibrium if $d_1\neq 0,1.$
We assume that in that case bidder $2$ believes that bidder $1$'s type is distributed by:
\begin{align}\label{beliefs}\mu(\cdot,d_1)=\begin{cases}
F(\cdot|d_1\leq v_1\leq  u),\quad &\text{ for }0\leq d_1\leq u,\\
F(\cdot|u\leq v_1\leq  \min\{1,d_1\}),\quad &\text{ for }u<d_1\leq \bar{d}.
\end{cases}\end{align}

Bidder $2$ thinks that bidder $1$ will bid $d_1$ for $d_1\leq u$. Therefore, bidder $2$, when observing $d_1\leq u$, will bid $d_1$ in case if this makes him positive profit, $v_2\geq d_1+cd_1$, and will not bid in the auction otherwise. Thus, the profit for bidder $1$ when depositing and bidding $d_1\leq u$ is as in the previous section:
$$F((1+c)d_1)v_1-cd_1.$$
This quantity should be non-positive for the marginal type:
\begin{align}F(u(1+c))\leq{c}.\end{align}

The expected profit for bidder $2$ of bidding $d_1\geq d_2\geq u$, if he observes $d_1>u$ is:
\begin{align*}&\Pi(v_2,d_1,d_2):=Pr[v_1\leq d_2|u\leq v_1\leq d_1]\left(v_2-E\left[v_1|u\leq  v_1\leq d_2\right]\right)-cd_2.
\end{align*}

Let $v(d_1)$ be the smallest valuation of bidder $2$ for which he enters the auction if he observes a deposit of $d_1>u$, i.e.$$v(d_1):=\min\{0\leq v_2\leq 1: \max_{u\leq d_2\leq d_1}\Pi(v_2,d_1,d_2)\geq 0\},$$
respectively $ v(d_1):=1$ if the min does not exist. We let $v:=v(1)$.

At the marginal type $v_1=u$, bidder $1$ must be indifferent between depositing $0$ and depositing $1$ and bidding his type. Note, moreover, that the marginal type when depositing $1$ and bidding his type either gets the object for free or does not get the object  (since bidder $2$ will either not bid at all in that case, $b_2=0$, or bid $b_2\geq u$, since he believes that bidder $1$ has a valuation of $u$ or higher). Thus, his payoff of depositing $1$ and bidding his type is
$$F(v)u-c.$$
The marginal type is indifferent between depositing $0$ and depositing $1$ if \begin{align}F(v)u=c.\end{align}

\subsubsection*{Quadratic Distribution}
For the quadratic distribution
\begin{equation}\label{Inequality_quadratic}(u(1+c))^2\leq c,\end{equation} and
\begin{equation}\label{Identity_quadratic}
uv^2=c.
\end{equation} 
which can be satisfied in general and pooling equilibria with two deposit levels exist:

\begin{proposition}
\label{prop:quadratic_pooling}
For large enough $c$, there exists a pooling equilibrium with two deposit levels.
\end{proposition}

The equilibrium exhibits the manifold inefficiencies we have highlighted in the introduction. Bidder~2 is deterred from entry unless his valuation is very high, see the Figure. Bidder~1 may win the auction even if he has a lower valuation. Bidder~1 engages in wasteful over-deposits in equilibrium.

\begin{figure*}[t]\label{bidding2}
  \centering
  \subfigure[Bidding (yellow) and deposit (blue) function for bidder $1$ for sequential depositing.]{\includegraphics[width=0.4\textwidth]{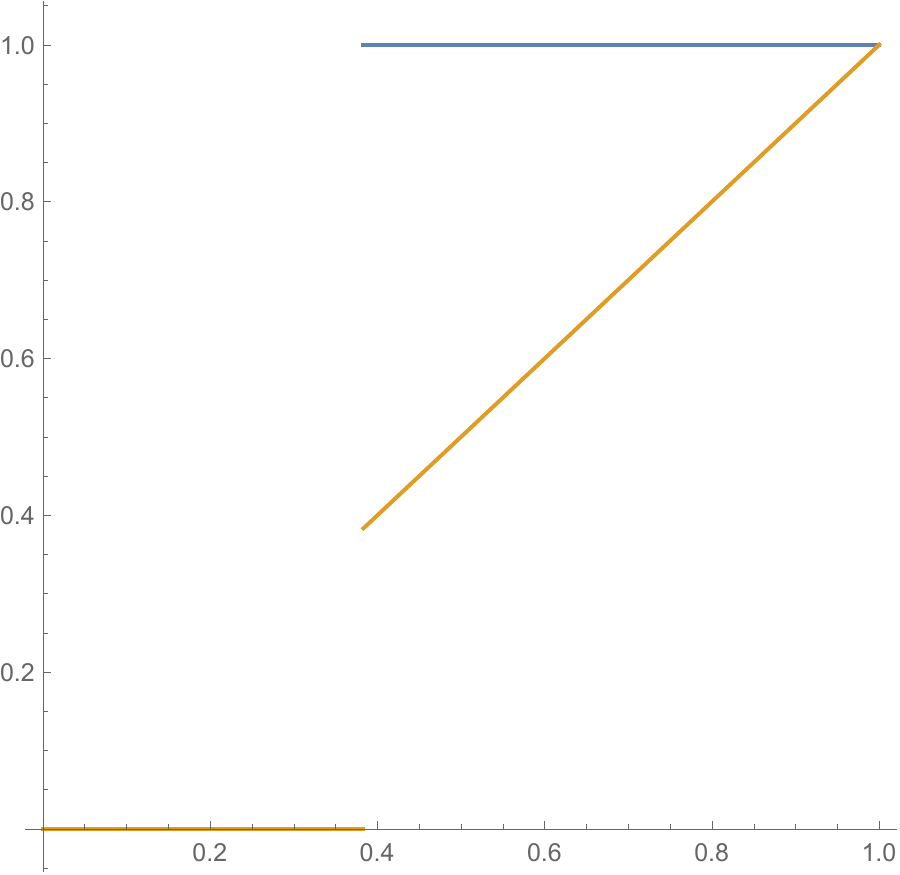}}
  \hspace{1.5cm}
  \subfigure[The bidding and depositing function for bidder $2$ for sequential depositing.]{\includegraphics[width=0.4\textwidth]{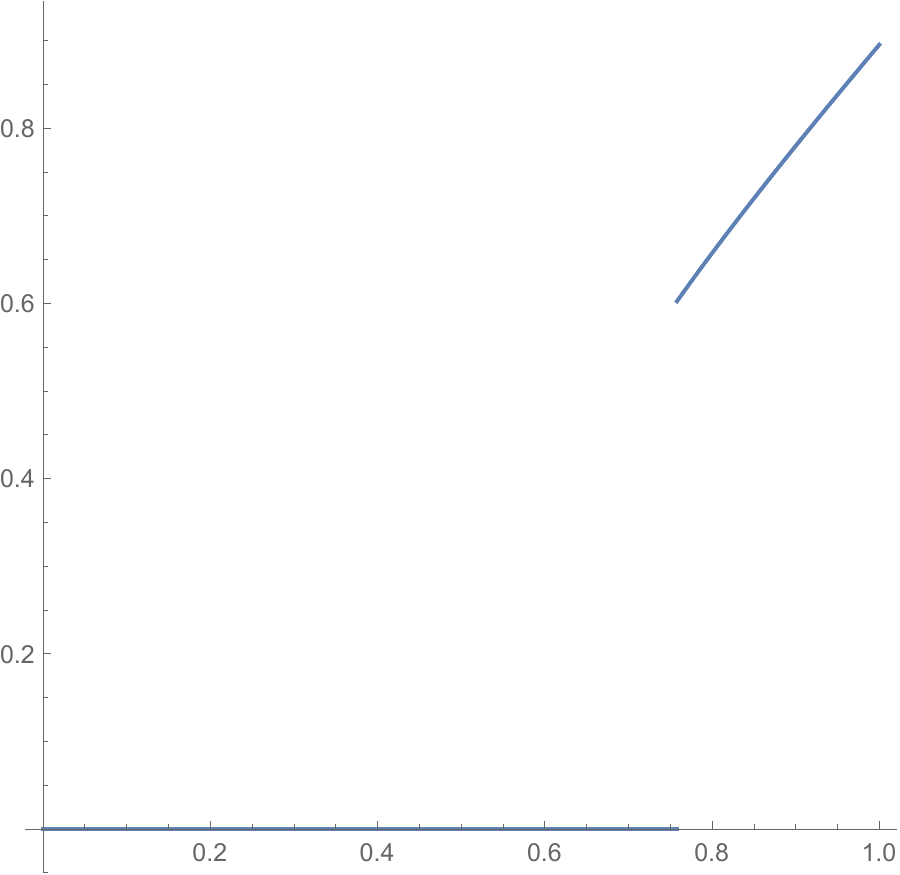}}
\end{figure*}





\section{Conclusion}

Given the potentially severe inefficiency of the auction design, it is a natural question how to improve it.
Several potential remedies, both cryptographic and economic, can be considered: Encryption of deposit levels is a natural idea. There are several blockchain projects with built-in privacy functionality, e.g.~\cite{oasis}, that could be used for this purpose. Privacy preservation in on-chain auction specifically has been analyzed in~\cite{possible_solution}.   Lowering the cost of depositing, for example by shortening the staking time, lowering borrowing costs so that it becomes cheaper to make large deposits or decreasing transaction costs can mitigate the problem. A more radical approach would  allow for under-collateralizing of bids where the residual sum has to be paid by the winning bidder only after winning. Finally it is a natural question whether other payment rules than the second price rule would allow for more efficient outcomes in the presence of deposits.
\bibliographystyle{plain}
\bibliography{sample}

\begin{thebibliography}{10}

\bibitem{all_pay_auctions}
Michael~R. Baye, Dan Kovenock, and Casper~G. de~Vries.
\newblock The all-pay auction with complete information.
\newblock {\em Economic Theory}, 8:291--305, 1996.

\bibitem{possible_solution}
Erik{-}Oliver Blass and Florian Kerschbaum.
\newblock {BOREALIS:} building block for sealed bid auctions on blockchains.
\newblock In Hung{-}Min Sun, Shiuh{-}Pyng Shieh, Guofei Gu, and Giuseppe
  Ateniese, editors, {\em {ASIA} {CCS} '20: The 15th {ACM} Asia Conference on
  Computer and Communications Security, Taipei, Taiwan, October 5-9, 2020},
  pages 558--571. {ACM}, 2020.

\bibitem{seq_costly_entry}
Xiaogang Che and Tilman Klumpp.
\newblock Entry deterrence in dynamic second-price auctions.
\newblock {\em American Economic Journal: Microeconomics}, 8(2):168--201, 2016.

\bibitem{cho1987}
In-Koo Cho and David~M Kreps.
\newblock Signaling games and stable equilibria.
\newblock {\em The Quarterly Journal of Economics}, 102(2):179--221, 1987.

\bibitem{flash_loan}
Philip Daian, Steven Goldfeder, Tyler Kell, Yunqi Li, Xueyuan Zhao, Iddo
  Bentov, Lorenz Breidenbach, and Ari Juels.
\newblock Flash boys 2.0: Frontrunning in decentralized exchanges, miner
  extractable value, and consensus instability.
\newblock In {\em 2020 {IEEE} Symposium on Security and Privacy, {SP} 2020, San
  Francisco, CA, USA, May 18-21, 2020}, pages 910--927. {IEEE}, 2020.

\bibitem{fudenberg}
Drew Fudenberg and Jean Tirole.
\newblock {\em Game Theory}.
\newblock MIT Press, 1991.

\bibitem{resale}
Rod Garratt and Thomas Tr\"{o}ger.
\newblock Speculation in standard auctions with resale.
\newblock {\em Econometrica}, 74(3):753--769, 2006.

\bibitem{hafner2021}
Samuel H{\"a}fner and Alistair Stewart.
\newblock Blockchains, front-running, and candle auctions.
\newblock {\em Front-Running, and Candle Auctions (May 14, 2021)}, 2021.

\bibitem{money_burning}
Jason~D. Hartline and Tim Roughgarden.
\newblock Optimal mechanism design and money burning.
\newblock In Cynthia Dwork, editor, {\em Proceedings of the 40th Annual {ACM}
  Symposium on Theory of Computing, Victoria, British Columbia, Canada, May
  17-20, 2008}, pages 75--84. {ACM}, 2008.

\bibitem{horner2007}
Johannes H{\"o}rner and Nicolas Sahuguet.
\newblock Costly signalling in auctions.
\newblock {\em The Review of Economic Studies}, 74(1):173--206, 2007.

\bibitem{mamageishvili2020}
Akaki Mamageishvili and Jan~Christoph Schlegel.
\newblock Optimal smart contracts with costly verification.
\newblock In {\em 2020 IEEE International Conference on Blockchain and
  Cryptocurrency (ICBC)}, pages 1--8. IEEE, 2020.

\bibitem{roth2002}
Alvin~E Roth and Axel Ockenfels.
\newblock Last-minute bidding and the rules for ending second-price auctions:
  Evidence from ebay and amazon auctions on the internet.
\newblock {\em American Economic Review}, 92(4):1093--1103, 2002.

\bibitem{signalling}
Michael Spence.
\newblock Job market signaling.
\newblock {\em The Quarterly Journal of Economics}, 87:355--374, 1973.

\bibitem{VCG}
William Vickrey.
\newblock Counterspeculation, auctions, and competitive sealed tenders.
\newblock {\em The Journal of Finance}, 16(1):8--37, 1961.

\bibitem{oasis}
Whitepaper.
\newblock The oasis blockchain protocol, 2020.

\end{thebibliography}

\newpage

\appendix

\section{Simultaneous depositing}\label{Simultanous}

We first consider the benchmark case of simultaneous bidding (and depositing). The following analysis works for any tie-breaking rule to determine the winner in case of equal bids. 
Since bids are simultaneous it is optimal not to over-deposit, that is $b(v_i)=d(v_i)$ (if $b(v_i)<d(v_i)$, then the bidder could deposit $b(v_i)$ instead, without changing his bid and become strictly better off by saving depositing cost).

Thus the equilibrium bidding function is fully determined by the equilibrium deposit function. We consider symmetric equilibria. Let $d:[0,1] \rightarrow[0,\bar{d}]$ be the equilibrium deposit function for both bidders. The expected profit for bidder $i$ when he deposits and bids an amount $d(w)$ corresponding to  type $w$ is given by
$$F(w)(v_i-E[d(v_j)|v_j\leq w])-cd(w),$$

Taking first order conditions and assuming that $d$ is differentiable, and using the fact that $d(0)=0$:
$$f(w)(v_i-E[d(v_j)|v_j\leq w])-F(w)f_{(0,w)}(w)d(w)-cd'(w)=0$$
where $f_{(0,w)}$ is the density conditional on the valuation being smaller $w$.
By the revelation principle the expression is optimized if he deposits and bids the amount $d(v_i)$ corresponding to his true valuation, i.e. when $w=v_i$, and therefore:
$$f(v_i)(v_i-E[d(v_j)|v_j\leq v_i])-F(v_i)f_{(0,v_i)}(v_i)d(v_i)-cd'(v_i)=0.$$

As an example we consider the case of an uniform distribution:
\subsubsection*{Uniform Distribution}
If valuations are uniformly distributed on $(0,1)$ the previous first order condition becomes:
$$v_i-d(v_i)-cd'(v_i)=0.$$

We can use the boundary condition $d(0)=0$ and obtain the solution to the differential equation:
$$d(v_i)=v_i-c(1-e^{-v_i/c}).$$



\section{Proofs}

\begin{proof}[Proof of Proposition~\ref{separationconditionalonentry}]
We define a threshold, $$\underline{u}:=\inf_{0\leq v\leq \tfrac{1}{1+c}}\{\max_{0\leq d\leq v}(vF((1+c)d)-cd)> 0\}.$$
For an equilibrium with (essential) separation conditional on entry, entering can only be profitable for types $v\geq\underline{u}$. 

Next we derive equilibrium depositing strategies for types $\underline{u}\leq v_1\leq \tfrac{1}{1+c}$. In the following we consider separating equilibrium strategies conditional on entry, i.e. for $\underline{u}\leq v_1,u_1\leq \tfrac{1}{1+c}$ with $u_1\neq v_1$ we require $d_1(v_1)\neq d_1(u_1)$.
We distinguish between two cases:
\begin{enumerate}
\item $(1+c)\underline{u}f((1+c)\underline{u})\geq c$
\item $(1+c)\underline{u}f((1+c)\underline{u})<c$
\end{enumerate}

In the first case, note that 
$(1+c)\underline{u}f((1+c)\underline{u})\geq c$ implies that the profit function for bidder $1$ for type $\underline{u}$ in case of under-depositing
$$\underline{u}F((1+c)d)-cd$$
has a non-negative derivative at $\underline{u}$ and therefore has a maximum at $d=\underline{u}$. Moreover, by continuity and the definition of $\underline{u}$ we have $\underline{u}F((1+c)d)-c\underline{u}=0$ and therefore $\underline{u}=\tfrac{1}{1+c}F^{-1}(c)$. More generally, for any valuation $v_1$ such that $(1+c)v_1f((1+c)v_1)\geq c$ the optimal strategy in case of under-depositing is to deposit $d_1=v_1$. Thus over-depositing is optimal if $1+c)v_1f((1+c)v_1)\geq c$. Over-depositing strategies satisfy the first order condition $$c d'(v_1)=(1+c)v_1f((1+c)v_1).$$
Integrating the first order condition:
$$d(v_1)-d(\underline{u})=\tfrac{1}{c}\int_{\underline{u}}^{v_1}(1+c)vf((1+c)v)dv=\frac{F((1+c)v_1)-F((1+c)\underline{u})}{c(1+c)}E[v|(1+c)\underline{u}\leq v\leq (1+c)v_1]$$
Using the assumption that we classify a maximally efficient equilibrium and under-depositing at $\underline{u}$ is not feasible, we can use the boundary condition $d(\underline{u})=\underline{u}$
to obtain:
$$d_1(v_1)=  \underline{u}+\frac{F((1+c)v_1)-c}{c(1+c)}E[v|(1+c)\underline{u}\leq v\leq (1+c)v_1]$$
We have derived a depositing strategy on  $[\underline{u},\bar{u}]$ for $\bar{u}=\inf\{\underline{u}\leq v_1\leq \tfrac{1}{1+c}:(1+c)v_1f((1+c)v_1)<c \}$ resp. $\bar{u}=\tfrac{1}{1+c}$ in case $(1+c)v_1f((1+c)v_1)\geq c $ for all $\underline{u}\leq v_1\leq\tfrac{1}{1+c}$.

Finally, we show that the same depositing strategy is also an equilibrium depositing strategy for $\bar{u}\leq v_1\leq\tfrac{1}{1+c}$. First we show that $v_1\leq d_1(v_1)$. By uni-modality if $d_1(\tfrac{1}{1+c})\geq\tfrac{1}{1+c}$ then $d_1(v_1)\geq v_1$ for each $\bar{u}\leq v_1\leq\tfrac{1}{1+c}$. We have \begin{align*}d_1(\tfrac{1}{1+c})-\underline{u}&=\tfrac{1}{c(1+c)}\int_{(1+c)\underline{u}}^{1}vf(v)dv=\tfrac{1}{c(1+c)}(E[v_1]-Pr[v_1\leq(1+c)\underline{u}]E[v_1|v_1\leq(1+c)\underline{u}])\\&\geq \tfrac{1}{c(1+c)}(E[v_1]-c(1+c)\underline{u})=\tfrac{E[v_1]}{c(1+c)}-\underline{u}\geq \frac{1}{1+c}-\underline{u},\end{align*}
where the penultimate inequality uses the previous observation that $\underline{u}=\tfrac{1}{1+c}F^{-1}(c)$ and therefore $c=Pr[v_1\leq(1+c)\underline{u}]$. The last inequality follows from $E[v]\geq c$. Therefore $d_1(\tfrac{1}{1+c})\geq\tfrac{1}{1+c}$ and $d_1(v_1)\geq v_1$ for all $\underline{u}\leq v_1\leq \tfrac{1}{1+c}$. Given the deposit function $d_1(v_1)$, it is still optimal for types $\bar{u}\leq v_1\leq \tfrac{1}{1+c}$ to choose the deposit level corresponding to their type since
the first order condition $$cd'(v_1)=(1+c)v_1f((1+c)v_1)$$ characterizes a maximum of $$\max_{\underline{u}\leq v\leq \tfrac{1}{1+c}}v_1F((1+c)v)-cd_1(v).$$ Moreover, since $d_1(v_1)>v_1$ for all $\bar{u}\leq v_1<\tfrac{1}{1+c}$ and since depositing strategies need to be increasing in types, under-depositing cannot be part of an equilibrium where $d_1$ describes the depositing function on the interval $[\underline{u},\bar{u}]$. Thus $d_1$ describes the maximally efficient equilibrium depositing strategy on the entire interval $[\underline{u},\tfrac{1}{1+c}]$.   

We have obtained the following equilibrium: Bidder $1$ deposits
\begin{align*}
    &d_1(v_1)=\begin{cases} 
    \underline{u}+\frac{1-c}{c(1+c)}E[v|(1+c)\underline{u}\leq v],\quad & v_1\geq \frac{1}{1+c},\\
    \underline{u}+\frac{F((1+c)v_1)-c}{c(1+c)}E[v|(1+c)\underline{u}\leq v\leq (1+c)v_1],\quad & \frac{1}{1+c}>v_1\geq \underline{u},\\
    0,\quad & v_1<\underline{u}.
    \end{cases}    \end{align*}
    and bids 
    \begin{align*}
    & b_1(v_1)=\begin{cases}
    v_1,\quad & v_1\geq \underline{u},\\
    0,\quad & v_1<\underline{u}.
    \end{cases}
    \end{align*}
Bidder $2$'s belief when observing $\underline{u}\leq d_1\leq d_1(\tfrac{1}{1+c})$ is 
$$\mu_2(v,d_1)=\begin{cases}
1,\quad\text{ for }v\geq v_1(d_1),\\
0,\quad\text{ for }v< v_1(d_1),
\end{cases}$$
where $v_1(d)$ is the inverse of the depositing function
and his belief when observing $ d_1\geq d_1(\tfrac{1}{1+c})$ is $$\mu_2(\cdot,d_1)=F(\cdot|v_1\geq\frac{1}{1+c}).$$

    We also need to specify beliefs off-equilibrium when bidder $2$ observes a deposit between $0$ and $\underline{u}$. One possibility is to assume that bidder $2$ believes that bidder $1$ has type $\underline{u}$.
Thus his beliefs when observing $0\leq d_1\leq \underline{u}$ is 
$$\mu_2(v,d_1)=\begin{cases}
1,\quad\text{ for }v\geq \underline{u},\\
0,\quad\text{ for }v<\underline{u}.
\end{cases}$$
One can readily verify that given the beliefs it is not optimal to choose off-equilibrium deposit levels.

In the second case, we define a threshold $$s:=\inf\{\underline{u}\leq v_1\leq\tfrac{1}{1+c}:(1+c)vf((1+c)v)> c\}.$$

If bidder $1$ has a type $\underline{u}\leq v_1\leq s$
and deposits $d\leq v_1$ then his profit is
$$v_1F((1+c)d)-cd.$$
By assumption that $v_1\leq s$, the profit function has a negative first derivative at $d=v_1$. If $v_1\leq m$ this would mean that the profit is negative, and therefore $v_1<\underline{u}$ which is a contradiction. Thus $m\leq v_1$, and moreover, the maximum of the profit function needs to be in a critical point $d<v_1$. Moreover, the critical point corresponding to a maximum needs to satisfy $m\leq d<v_1$ (since critical points  $d_1<m$ correspond to local minima by convexity). The maximal point is therefore characterized by (the unique) $m\leq d<v_1$ that satisfies:
$$v_1f((1+c)d)=c,$$
and we have under-depositing in equilibrium. For types $v_1\geq s$ with the same logic as in case $1$ we obtain the depositing function:
$$d_1(v_1)=s+\frac{F((1+c)v_1)-F((1+c)s)}{c(1+c)}E[v|(1+c)s\leq v\leq (1+c)v_1].$$
The depositing function $d_1$ satisfies $d_1(v_1)\geq v_1$ for $s\leq v_1\leq \tfrac{1}{1+c}$ because  we have \begin{align*}d_1(\tfrac{1}{1+c})-s&=\tfrac{1}{c(1+c)}\int_{(1+c)s}^1vf(v)dv=\tfrac{1}{c(1+c)}(E[v_1]-\int_{0}^{(1+c)s}vf(v)dv)\\&\geq \tfrac{1}{c(1+c)}(E[v_1]-c(1+c)s)=\tfrac{E[v_1]}{c(1+c)}-s\geq \frac{1}{1+c}-s,\end{align*}
where the penultimate inequality follows as $(1+c)v_1f((1+c)v_1)\leq c$ for $v_1\leq s$ and the last inequality follows from $E[v]\geq c$. Therefore $d_1(\tfrac{1}{1+c})\geq\tfrac{1}{1+c}$ and by uni-modality $d_1(v_1)\geq v_1$ for all $s\leq v_1\leq \tfrac{1}{1+c}$. 

Thus we have an equilibrium where bidder $1$  under-deposits for types $v_1\leq s$ and over-deposits for types $s\leq v_1$. The equilibrium strategies are as follows: Bidder $1$ deposits
\begin{align*}
    &d_1(v_1)=\begin{cases} 
    s+\frac{1-F((1+c)s)}{c(1+c)}E[v|(1+c)s\leq v],\quad & v_1\geq \frac{1}{1+c},\\
    s+\frac{F((1+c)v_1)-F((1+c)s)}{c(1+c)}E[v|(1+c)s\leq v\leq (1+c)v_1],\quad & \frac{1}{1+c}>v_1\geq s,\\
        \tfrac{1}{1+c}f^{-1}\left(\tfrac{c}{(1+c)v}\right),\quad & s> v_1\geq \underline{u},\\
    0,\quad & v_1<\underline{u}.
    \end{cases}    \end{align*}
    and bids 
    \begin{align*}
    & b_1(v_1)=\begin{cases}
    v_1,\quad & v_1\geq (1+c)s,\\
    d_1,\quad & s>v_1\geq \underline{u},\\
    0,\quad & v_1<\underline{u}.
    \end{cases}
    \end{align*}
Bidder $2$'s belief when observing $\underline{u}\leq d_1\leq d_1(\tfrac{1}{1+c})$ is 
$$\mu_2(v,d_1)=\begin{cases}
1,\quad\text{ for }v\geq v_1(d_1),\\
0,\quad\text{ for }v< v_1(d_1),
\end{cases}$$
where $v_1(d)$ is the inverse of the depositing function
and his belief when observing $ d_1\geq d_1(\tfrac{1}{1+c})$ is $$\mu_2(\cdot,d_1)=F(\cdot|v_1\geq\frac{1}{1+c}).$$
We also need to specify beliefs off-equilibrium when bidder $2$ observes a deposit between $0$ and $\underline{u}$. One possibility is to assume that bidder $2$ believes that bidder $1$ has type $\underline{u}$.
Thus his beliefs when observing $0\leq d_1\leq \underline{u}$ is 
$$\mu_2(v,d_1)=\begin{cases}
1,\quad\text{ for }v\geq \underline{u},\\
0,\quad\text{ for }v<\underline{u}.
\end{cases}$$

In conclusion, there is an equilibrium that is essentially (except for the very highest types with valuation above $\frac{1}{1+c}$) separating conditional on entry.
\end{proof}

\begin{proof}[Proof of Corollary~\ref{uniform_separation}]
In case of under-depositing the profit is
$$v_1F((1+c)d_1)-cd_1,$$
which is positive only if $v_1\geq \underline{u}:=\frac{F^{-1}(c)}{1+c}$. Thus, types  $0\leq v_1\leq \underline{u}$ over-deposit or do not enter the auction at all. In case of over-depositing Proposition~\ref{separationconditionalonentry} characterizes an equilibrium. Note that the lower bound requirement on $E[v_1]$ is now redundant because convexity implies $(1+c)v_1f((1+c)v_1)\geq c$ for each $v_1\geq\underline{u}$.
 \end{proof}

\begin{proof}[of Proposition~\ref{prop:quadratic_pooling}]
First, we determine the bidding function for bidder $2$, if he observes $d_1>u$ conditional on entering the auction. For $u<d_2\leq d_1$, we have 

\begin{align*}
&\Pi(v_2,d_1,d_2)=\frac{d_2^2-u^2}{d_1^2-u^2}(v_2-\int_{u}^{d_2}v_1\frac{2v_1}{d_2^2-u^2}dv_1)-cd_2=\frac{d_2^2-u^2}{d_1^2-u^2}v_2-\frac{2(d_2^3-u^3)}{3(d_1^2-u^2)}-cd_2.
\end{align*}

Taking first order conditions:
$$\frac{\partial \Pi}{\partial d_2}=\frac{2d_2v_2}{d_1^2-u^2}-\frac{2d_2^2}{d_1^2-u^2}-c=0.$$
There are two critical points of which the larger one corresponds to a local maximum:

\begin{equation}\label{second_bidder_deposit_quadratic}
d_2=\frac{1}{2}\left(v_2+\sqrt{v_2^2-2c(d_1^2-u^2)}\right).
\end{equation}

To determine the marginal types $u$ and $v$ we need to check whether there are $$0<u<v\leq 1$$
such that equations~\eqref{Identity_quadratic} and~\eqref{second_bidder_deposit_quadratic} with $d_1=1$ and $v_2=v$, inequality~\eqref{Inequality_quadratic} are satisfied and
\begin{align}
  (d_2^2-u^2)v-\frac{2}{3}(d^3_2-u^3)=cd_2(1-u^2)  
\end{align}
holds (i.e.,~the profit for the marginal type of bidder $2$ is equal to $0$).
Similarly, to determine $v(d_1)$, we need to find a $v(d_1)$ that satisfies Equation~\eqref{second_bidder_deposit_quadratic} with $v_2=v(d_1)$ and 

\begin{align}\label{quadratic_pooling}
  (d_2^2-u^2)v(d_1)-\frac{2}{3}(d^3_2-u^3)=cd_2(d_1^2-u^2).
\end{align}

The equilibrium is as follows. Bidder $1$ deposits 
$$d_1(v_1)=\begin{cases} 1,\quad & v_1\geq u,\\
    0,\quad & v_1<u,
    \end{cases}$$ and bids
\begin{align*}
b_1(v_1)=\begin{cases} v_1,\quad & v_1\geq u,\\
    0,\quad & v_1<u,
    \end{cases}
    \end{align*}
If bidder $2$ observes a deposit $d_1>u$ above the threshold he deposits
        \begin{align*}
    &d_2(v_2,d_1)=\begin{cases} \min\left\{d_1,\frac{1}{2}\left(v_2+\sqrt{v_2^2-2c(d_1^2-u^2)}\right)\right\},\quad & v_2\geq v(d_1),\\
    0,\quad & v_2<v(d_1),
    \end{cases}\end{align*}
and if he observes a deposit $d_1\leq u$ below the threshold he deposits
\begin{align*}
    &d_2(v_2,d_1)=\begin{cases} d_1,\quad & v_2\geq d_1,\\
    0,\quad & v_2<d_1.
    \end{cases}
\end{align*}
In either case he bids his deposit, $b_2(v_2,d_1)=d_2(v_2,d_1)$. His beliefs are given by Equation~\ref{beliefs}.
Next, we verify that the deposit functions derived above with the beliefs specified above form a Perfect Bayesian Equilibrium if valuations are quadratically distributed for the case that the marginal cost is $c=0.22$. In this case the marginal type is $u=0.382981$ and bidder $2$ will enter the auction only if his valuation is above $v=0.757919$. Note that incentive compatibility for the marginal type of bidder $1$ holds as $$u^2(1+c)^2=0.382981^2\cdot1.22^2\approx0.21831<c= 0.22.$$ It remains to show that the highest type of bidder $1$ and the marginal type of bidder $1$ cannot profit by depositing $u<d_1<1$ (it is not profitable to deposit more than $1$ since depositing becomes more costly but the bidding behaviour of bidder $2$ does not change). It is straightforward then to  see that also the types $u<v_1<1$ cannot profit by depositing $u<v_1<1$. The expected profit of bidder $1$ when depositing $d_1<1$ is
  \begin{align*} &Pr[d_1\geq d_2(d_1,v_2)]-Pr[d_1\geq d_2(d_1,v_2)]E[d_2(d_1,v_2)|d_1\geq d_2(d_1,v_2)]-cd_1.
  \end{align*}
  Denote by $\bar{v}(d_1)$ the type of bidder $2$ with $d_1=d_2(d_1,\bar{v}(d_1))$. We have:
  $$d_1=\frac{1}{2}(\bar{v}(d_1)+\sqrt{\bar{v}(d_1)^2-2c(d_1^2-u^2)})\Rightarrow \bar{v}(d_1)=d_1-\frac{2c(d_1^2-u^2)}{4d_1},$$
  and the profit of bidder $1$ when depositing $d_1$ is
  \begin{align*}
  &\bar{v}(d_1)^2-\int_{v(d_1)}^{\bar{v}(d_1)}2v_1d_2(d_1,v_2)dv_1-cd_1\leq \bar{v}(d_1)^2-\\ &\int_{v}^{\bar{v}(d_1)}2v_1d_2(d_1,v_2)dv_1-cd_1=\bar{v}(d_1)^2-\frac{1}{3} \\& \left(\bar{v}(d_1)^3-v^3+
  (\bar{v}(d_1)^2-2c(d_1^2-u^2))^{3/2}-(v^2-2c(d_1^2-u^2))^{3/2}\right)-cd_1,
  \end{align*}
  
  where the inequality follows as the marginal type $v(d_1)$ of bidder $2$ that enters the auction is decreasing in the deposit.
  For $c=0.22$, $u=0.382981$ and $v=0.757919$ this is an increasing function in $d_1$ for $d_1\geq u$. Thus, it is optimal for bidder $1$ to deposit the maximal amount $\bar{d}$ if his type is $v_1=1$.


Next, we consider the marginal type $v_1=u.$ Differentiating Equation~\eqref{quadratic_pooling} with respect to $d_1$ gives:
$$ (2d_2v(d_1)-2d_2^2)\frac{\partial d_2}{\partial d_1}+(d_2^2-u^2)\frac{\partial v(d_1)}{\partial d_1}=2cd_1d_2+c(d_1^2-u^2)\frac{\partial d_2}{\partial d_1}.$$

Therefore,
\begin{align*}
 & \frac{\partial v(d_1)}{\partial d_1}=
  \frac{1}{d_2^2-u^2}\left(2cd_1d_2-(c(d_1^2-u^2)-2d_2(v(d_1)-d_2))\frac{\partial d_2}{\partial d_1}\right)\\
  &=
  \frac{1}{d_2^2-u^2}\left(2cd_1d_2-(c(d_1^2-u^2)-2d_2\frac{1}{2}\left(v_2-\sqrt{v_2^2-2c(d_1^2-u^2)}\right)\frac{\partial d_2}{\partial d_1}\right)\\
   &=
  \frac{1}{d_2^2-u^2}\left(2cd_1d_2-(c(d_1^2-u^2)-\frac{1}{2}\left(v_2^2-(v_2^2-2c(d_1^2-u^2))\right)\frac{\partial d_2}{\partial d_1}\right)\\
  &=\frac{2cd_1d_2}{d_2^2-u^2}\geq\frac{2cd_1d_2}{d_1d_2-u^2}\geq\frac{2cd_2}{d_2-u^2/d_1}\geq \frac{2cv(d_1)}{v(d_1)-u^2/d_1},
  \end{align*} 
  where the first inequality follows as $d_2\leq d_1$ and the last inequality follows as $d_2\leq v(d_1)$ and the term $\frac{2cd_2}{d_2-u^2/d_1}$ is decreasing in $d_2$. 
   For the marginal type $v_1=u$, differentiating the profit as function of $d_1<\bar{d}$ with respect to $d_1$ yields
  \begin{align*}
      &\frac{\partial }{\partial d_1}(v(d_1)^2u-cd_1)=2v(d_1)u\frac{\partial v}{\partial d_1}-c\geq \frac{4cv(d_1)^2u}{v(d_1)-u^2/d_1}-c\geq\frac{4c(2u^2/d_1)^2u}{u^2/d_1}-c=16cu^3/d_1-c,
      \end{align*}
which is positive for $d_1\leq 0.89877$. On the other hand, we have $$\frac{2cd_1d_2}{d_2^2-u^2}\geq\frac{2cd_1d_2}{d_2^2}\geq\frac{2cd_1}{d_2}\geq \frac{2cd_1}{v(d_1)},$$
and, therefore
$$2v(d_1)u\frac{\partial v}{\partial d_1}-c\geq 4cd_1u-c=1.531924 cd_1-c$$
which is positive for $d_1>0.89877.$
Thus, bidder $1$'s profit is decreasing in $d_1$ if $v_1=u.$
\end{proof}
\begin{figure}[ht]
    \centering
    \includegraphics[width=16cm]{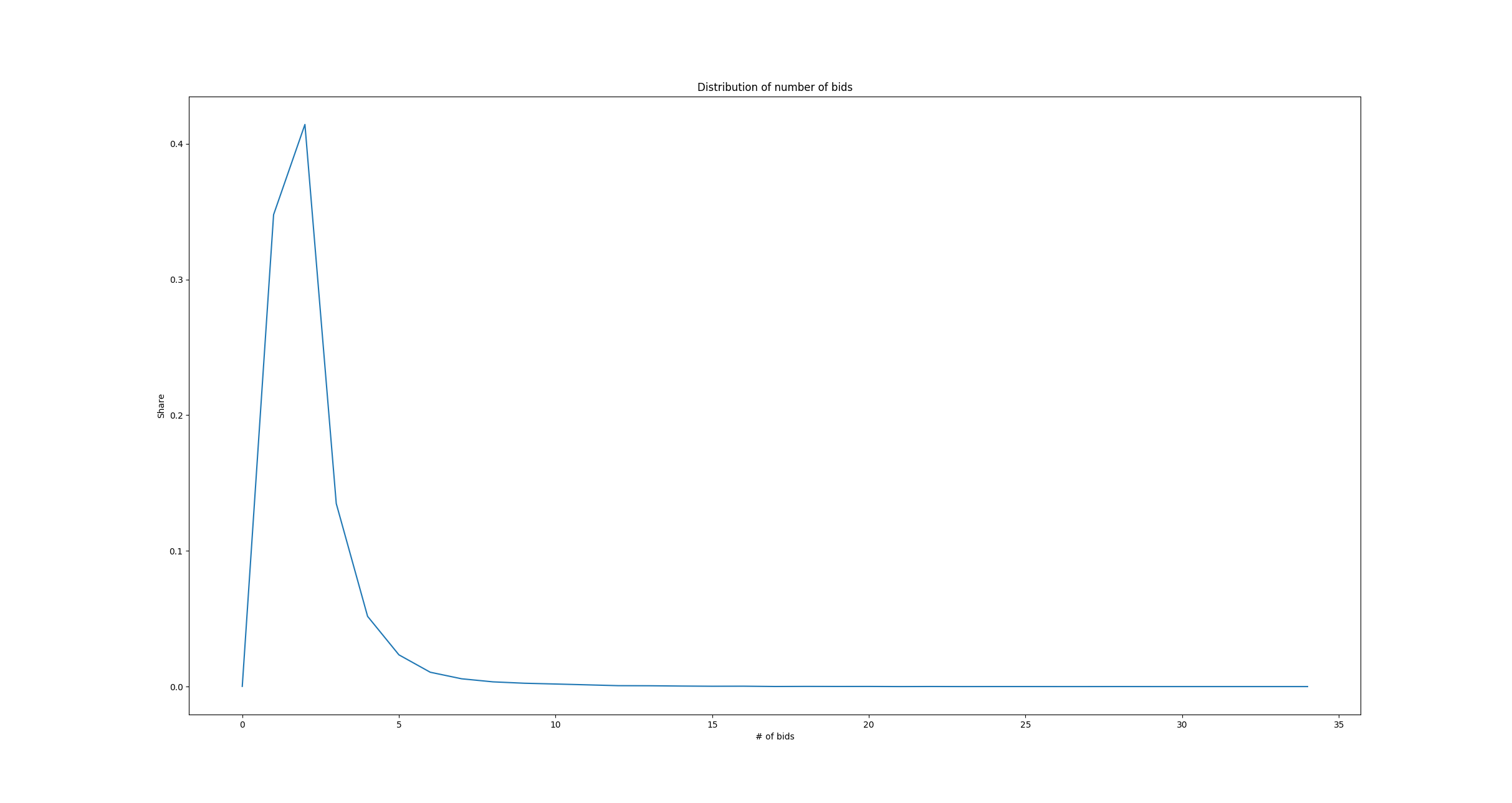}
    \caption{Distribution of number of bids}
    \label{fig:my_label}
\end{figure}
\begin{figure}[ht]
    \centering
    \includegraphics[width=16cm]{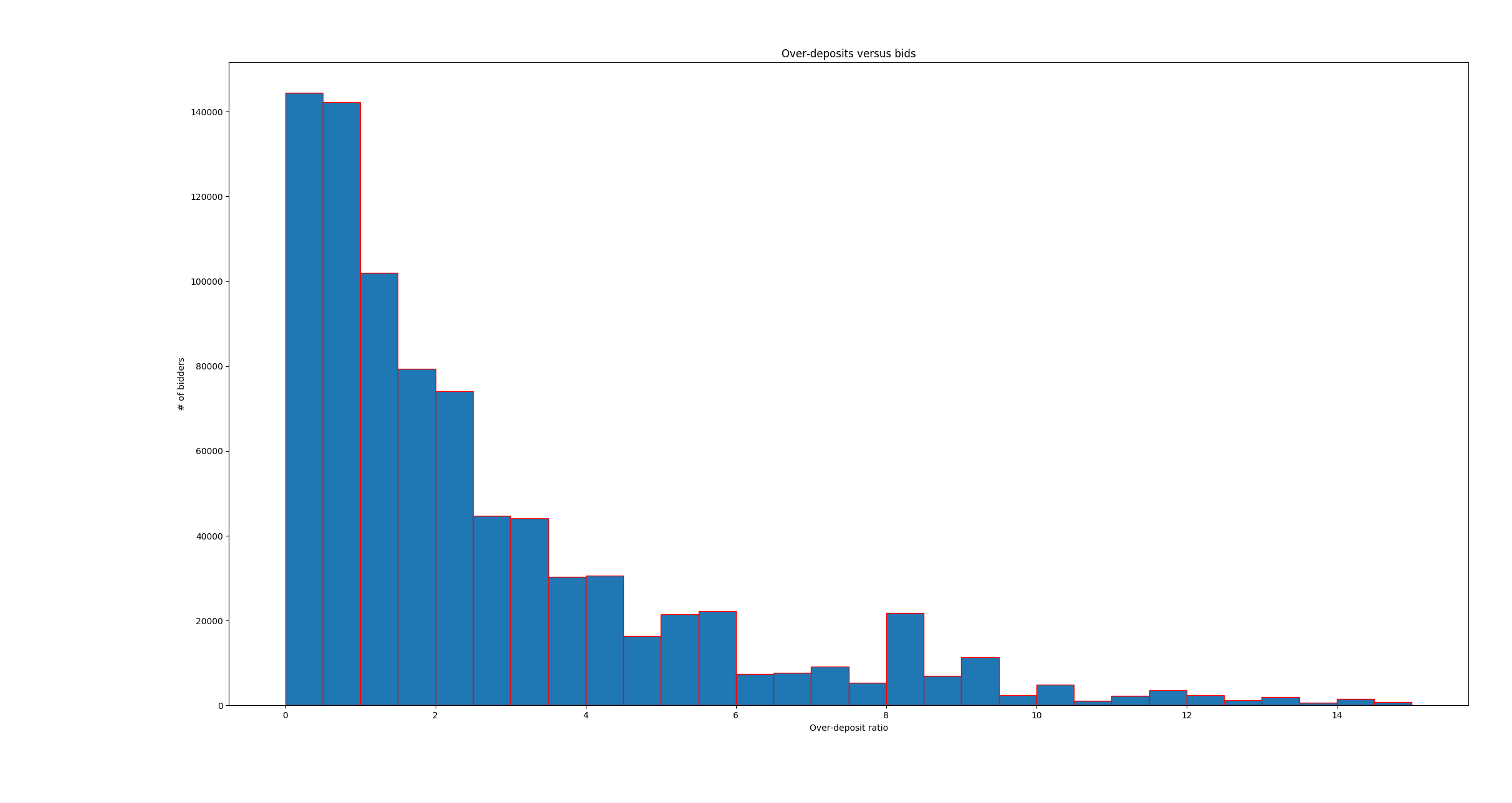}
    \caption{Overdepositing}
    \label{overdepositing}
\end{figure}
\end{document}